\documentclass[longauth]{aa}  

\usepackage{amsmath}
\usepackage{graphicx}
\usepackage{txfonts}
\usepackage{natbib}
\bibpunct{(}{)}{;}{a}{}{,} 
\begin{document}

   \title{Solar wind current sheets and deHoffmann-Teller analysis:}
   \subtitle{First results of DC electric field measurements by Solar Orbiter}
   \author{K. Steinvall
          \inst{1,2}
          \and
          Yu. V. Khotyaintsev\inst{1}
          \and
          G. Cozzani\inst{1}
          \and
          A. Vaivads\inst{3}
          \and
          E. Yordanova\inst{1}
          \and
          A. I. Eriksson\inst{1}
          \and
          N. J. T. Edberg\inst{1}
          \and
          M. Maksimovic\inst{4}
          \and
          S. D. Bale\inst{5,6}
          \and T. Chust\inst{7}
          \and V. Krasnoselskikh\inst{8,5}
          \and M. Kretzschmar\inst{8,9}
          \and E. Lorf\`evre\inst{10}
          \and D. Plettemeier\inst{11}
          \and J. Sou\v{c}ek\inst{12}
          \and M. Steller\inst{13}
          \and \v{S}. \v{S}tver\'ak\inst{14}
          \and A. Vecchio\inst{4,15}
          \and T. S. Horbury\inst{16}
          \and H. O'Brien\inst{16}
          \and V. Evans\inst{16}
          \and A. Fedorov\inst{17}
          \and P. Louarn\inst{17}
          \and V. G\'enot\inst{17}
          \and N. Andr\'e\inst{17}
          \and B. Lavraud\inst{17,18}
          \and A. P. Rouillard\inst{17}
          \and C. J. Owen\inst{19}
          }
   \institute{Swedish Institute of Space Physics (IRF), Uppsala 75121, Sweden\\
              \email{konrad.steinvall@irfu.se}
         \and
             Space and Plasma Physics, Department of Physics and Astronomy, Uppsala University, Uppsala 75120, Sweden
        \and
             Division of Space and Plasma Physics, School of Electrical Engineering and Computer Science,
KTH Royal Institute of Technology, Stockholm 11428, Sweden
\and
LESIA, Observatoire de Paris, Universit\'e PSL, CNRS, Sorbonne Universit\'e, Univ. Paris Diderot, Sorbonne Paris Cit\'e, 5 place Jules Janssen, 92195 Meudon, France
\and
Space Sciences Laboratory, University of California, Berkeley, CA, USA
\and
Physics Department, University of California, Berkeley, CA, USA
\and
LPP, CNRS, Ecole Polytechnique, Sorbonne Universit\'{e}, Observatoire de Paris, Universit\'{e} Paris-Saclay, Palaiseau, Paris, France
\and
LPC2E, CNRS, 3A avenue de la Recherche Scientifique, Orl\'eans, France
\and
Universit\'e d'Orl\'eans, Orl\'eans, France
\and
CNES, 18 Avenue Edouard Belin, 31400 Toulouse, France
\and
Technische Universität Dresden, Helmholtz Str. 10, D-01187 Dresden, Germany
\and
Institute of Atmospheric Physics of the Czech Academy of Sciences, Prague, Czechia
\and
Space Research Institute, Austrian Academy of Sciences, Graz, Austria
\and
Astronomical Institute of the Czech Academy of Sciences, Prague, Czechia
\and
Radboud Radio Lab, Department of Astrophysics, Radboud University, Nijmegen, The Netherlands
\and
Imperial College London, South Kensington Campus, London SW7 2AZ, UK
\and
Institut de Recherche en Astrophysique et Plan\'etologie, 9, Avenue du Colonel ROCHE, BP 4346, 31028 Toulouse Cedex 4, France
\and
Laboratoire d'astrophysique de Bordeaux, Univ. Bordeaux, CNRS, Pessac, France
\and
Mullard Space Science Laboratory, University College London, Holmbury St. Mary, Dorking, Surrey RH5 6NT, UK
\\
             }

   \date{}

  \abstract
   {Solar Orbiter was launched on February 10, 2020 with the purpose of investigating solar and heliospheric physics using a payload of instruments designed for both remote and in-situ sensing. Similar to the recently launched Parker Solar Probe, and unlike earlier missions, Solar Orbiter carries instruments designed to measure the low frequency DC electric fields.}  
   {In this paper we assess the quality of the low-frequency DC electric field measured by the Radio and Plasma Waves instrument (RPW) on Solar Orbiter. In particular we investigate the possibility of using Solar Orbiter's DC electric and magnetic field data to estimate the solar wind speed.}
   {We use deHoffmann-Teller (HT) analysis based on measurements of the electric and magnetic fields to find the velocity of solar wind current sheets which minimizes a single component of the electric field. By comparing the HT velocity to proton velocity measured by the Proton and Alpha particle Sensor (PAS) we develop a simple model for the effective antenna length, $L_\text{eff}$ of the E-field probes. We then use the HT method to estimate the speed of the solar wind.}
   {Using the HT method, we find that the observed variations in $E_y$ are often in excellent agreement with the variations in the magnetic field. The magnitude of $E_y$, however, is uncertain due to the fact that the $L_\text{eff}$ depends on the plasma environment. We derive an empirical model relating $L_\text{eff}$ to the Debye length, which we can use to improve the estimate of $E_y$ and consequently the estimated solar wind speed.}
   {The low frequency electric field provided by RPW is of high quality. Using deHoffmann-Teller analysis, Solar Orbiter's magnetic and electric field measurements can be used to estimate the solar wind speed when plasma data is unavailable.}

 \titlerunning{Solar wind current sheets and deHoffmann-Teller analysis}
   \maketitle
%

\section{Introduction}

With the recent launch of NASA's Parker Solar Probe~\citep{Fox2016} and ESA's Solar Orbiter~\citep{Muller2020}, the low-frequency 'DC' electric field $\mathbf{E}$, associated with the solar wind can be measured for the first time at heliocentric distances below 1 AU. The electric field is one of the most challenging quantities to measure. One significant complication is that the spacecraft and its solar panels are charged to some variable potential and can generate an electrostatic field which is not of physical interest~\citep{Cully2007,Johansson2020}. Another difficulty is that the E-field probes must be adequately separated to provide the sensitivity necessary to measure weak electric fields in plasma. To investigate the quality of the E-field measurement, it can often be useful to analyse large-scale fluctuations in the magnetic field $\mathbf{B}$. Since the solar wind plasma is on large scales in the ideal MHD state, $\mathbf{E}$ is related to $\mathbf{B}$ via the simple formula $\mathbf{E}=-\mathbf{v}\times\mathbf{B}$, where $\mathbf{v}$ is the plasma bulk velocity. Thus, if a magnetic structure such as a current sheet crosses the spacecraft at a constant $\mathbf{v}$, $\mathbf{E}$ fluctuations linearly related to the fluctuations in $\mathbf{B}$ will be observed. Moreover, if the plasma velocity is measured independently one can calculate the electric field as $\mathbf{E}=-\mathbf{v}\times\mathbf{B}$ and compare the results directly to the measured E-field as a quality control~\citep{Mozer2020b}.

When analysing magnetic structures such as current sheets, Alfv\'en waves, or shocks, it is often necessary to find the proper frame of the structure. This is commonly done using deHoffmann-Teller analysis~\citep{deHoffmann1950,Sonnerup1987}. The proper frame of a magnetic structure is a deHoffmann-Teller (HT) frame if $\mathbf{E}=0$ in it. Thus, if the HT frame exists, the electric field in the spacecraft frame is given by $\mathbf{E}=-\mathbf{v}_\text{HT}\times\mathbf{B}$, where the HT velocity $\mathbf{v}_\text{HT}$, is the velocity of the HT frame with respect to the spacecraft frame. We note that we cannot transform away any component of $\mathbf{E}$ parallel to $\mathbf{B}$ ($E_\parallel$). However, on the large scales we are interested in, $E_\parallel$ is typically zero. HT analysis has been used extensively to analyse shocks~\citep[e.g.][]{Schwartz1988,Lefebvre2007} and magnetic reconnection at Earth's magnetopause~\citep[e.g.][]{Fuselier1991,Phan2004} and in the solar wind~\citep[e.g.][]{Gosling2005}. More recently,~\citet{Horbury2020b} used HT analysis to find the proper frame of large-scale solar wind spikes~\citep{Horbury2018}, also known as 'switchbacks'~\citep[e.g.][]{Bale2019,Mozer2020}, observed by Parker Solar Probe, and \citet{Nemecek2020} applied HT analysis on particle velocity data in the solar wind, concluding that the HT frame can be considered a proper solar wind frame.

Another important application of HT analysis, which to the best of our knowledge has not been previously reported, is to estimate the solar wind speed in the absence of particle data. Fluctuations in the solar wind magnetic field are primarily either frozen-in current sheets or MHD turbulence~\citep{Borovsky2010}. The HT frame of a current sheet is the frame in which $\mathbf{E}=0$ on both sides of the discontinuity. In other words, it is the frame in which the plasma flow is field aligned on both sides. In the solar wind the plasma flow is usually approximately constant across the discontinuity, while $\mathbf{B}$ is arbitrarily rotated in the plane of the discontinuity, implying $\mathbf{v}$ cannot be field-aligned simultaneously on both sides (unless the fields are exactly parallel or anti-parallel). It follows that if a HT frame exists, $\mathbf{v}$ must be zero in it, implying $\mathbf{v}_\text{HT}=\mathbf{v}$. Thus, by finding $\mathbf{v}_\text{HT}$ of solar wind current sheets, which are often tangential discontinuities~\citep{Knetter2004}, we find a measure of the solar wind velocity. For completeness we note that while rotational discontinuities always have a HT frame, there is a specific theoretical configuration in which tangential discontinuities do not have a HT frame. This special configuration has perfectly parallel or anti-parallel magnetic fields on the two sides of the discontinuity, while the plasma flow perpendicular to the two magnetic fields differ~\citep{khrabrov1998}. However, due to the strict criteria on the magnetic fields, this situation does not occur in practice, and the discontinuities we observe in the solar wind should always have a HT frame.

If we instead apply HT analysis on an Alfv\'enic structure, Faraday's law implies that $\mathbf{v}_\text{HT}$ is the phase velocity of the structure. In the spacecraft frame this gives $\mathbf{v}_\text{HT}=\mathbf{v}+\mathbf{v}_\phi$, where $\mathbf{v}_\phi$ is the phase velocity in the plasma frame, which depends on the wave-mode and propagation direction, and typically is of the order of the Alfv\'en speed $v_A$. Far from the Sun $v_A$ is small compared to $\mathbf{v}$, and we can treat it as a small correction to the solar wind speed. However, closer to the Sun this extra contribution might become non-negligible when analysing Alfv\'enic structures such as spikes, as discussed by \citet{Horbury2020b}.

To avoid potential confusion, we want to emphasize that while the HT velocity is derived from $\mathbf{E}=-\mathbf{v}\times\mathbf{B}$, $v_\text{HT}$ is not necessarily perpendicular to the ambient magnetic field. One way to understand this is to realize that we are applying the analysis to magnetic structures which have associated magnetic fluctuations $\delta\mathbf{B}$ ($\mathbf{B} = \mathbf{B}_0 + \delta\mathbf{B}$). For simplicity, if $\mathbf{v}$ is constant, the corresponding electric fluctuation is given by
\begin{equation}
    \delta\mathbf{E}=-\mathbf{v}\times\delta\mathbf{B}.
\end{equation}
So even if $\mathbf{B}_0$ and $\mathbf{v}$ are parallel, as long as $\delta\mathbf{B}$ has a component perpendicular to $\mathbf{B}_0$, it can be used together with $\delta\mathbf{E}$ to estimate $\mathbf{v}$.

In this study we apply HT analysis based on electric and magnetic field measurements on solar wind current sheets to investigate the quality of Solar Orbiter's DC electric field. Once we have established that the quality is good, we apply the HT analysis to estimate the solar wind speed. We find that our method can be used to distinguish between fast and slow solar wind, providing us with a measure of the solar wind velocity even when particle data is unavailable.

\section{Instrumentation and DC $\mathbf{E}$ calibration}
\label{sec:cal}
\begin{figure}
\centering
\resizebox{\hsize}{!}{\includegraphics{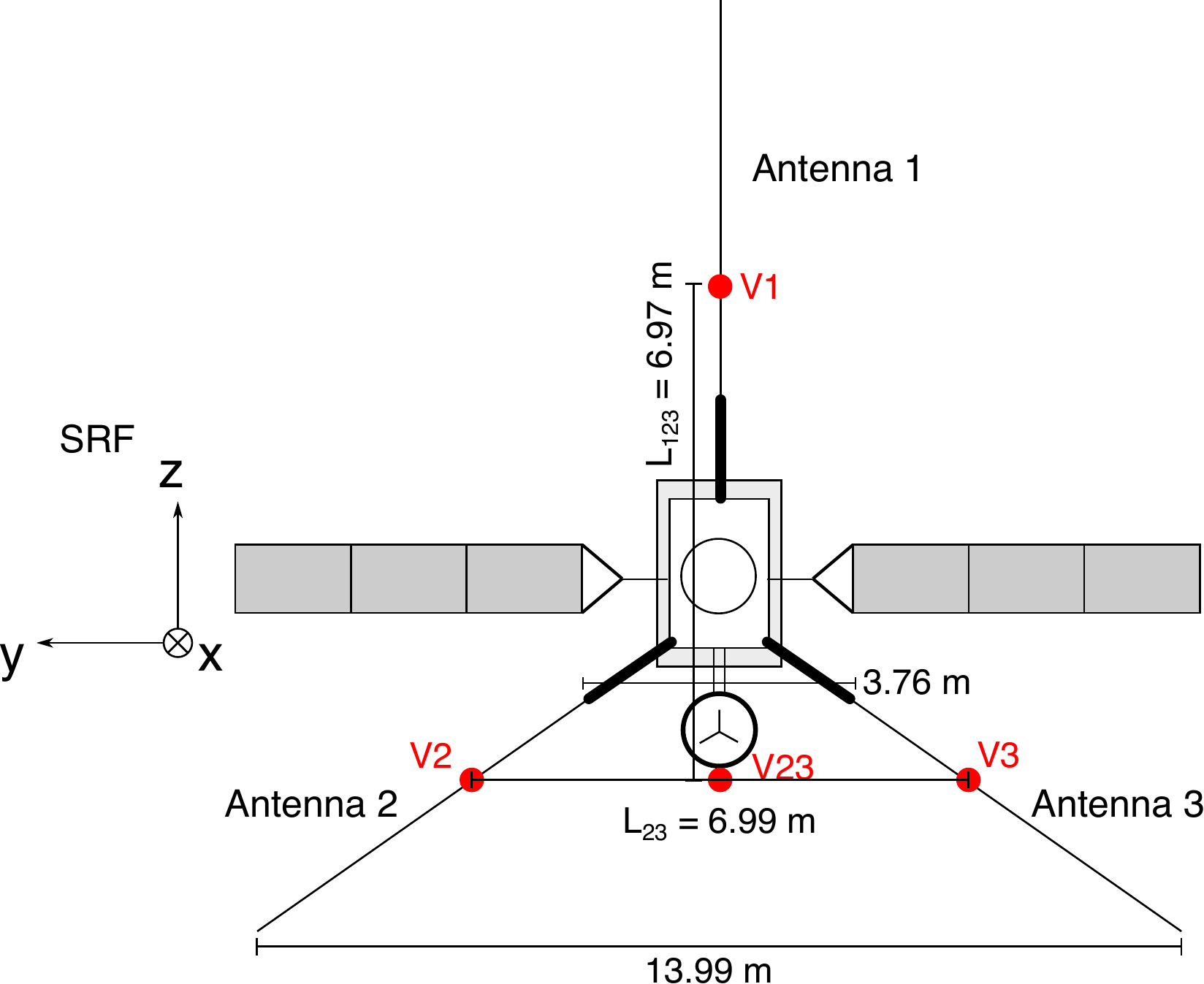}}
\caption{Qualitative sketch of Solar Orbiter as seen from behind, showing the RPW antennas (i.e. the probes) and the nominal effective antenna lengths $L_{23}$ and $L_{123}$.}
     \label{fig:solo}
\end{figure}
The data used in this study are from the Solar Orbiter mission. We use magnetic field data from the magnetometer (MAG)~\citep{Horbury2020} sampled at 8 vectors/second, plasma measurements from the Solar Wind Analyser suite (SWA)~\citep{Owen2020}, in particular from the Proton and Alpha particle Sensor (PAS) at 0.25 samples/second, and electric fields from the Radio and Plasma Waves instrument (RPW)~\citep{Maksimovic2020} sampled at 16 vectors/second.
Throughout this Paper, we primarily present vector quantities in the spacecraft coordinate system (SRF) defined in such a way that $\hat{\mathbf{x}}$ is pointing from Solar Orbiter toward the Sun, $\hat{\mathbf{z}}$ is along RPW antenna 1 (approximately normal to the ecliptic plane for the data used in this study, see Fig.~\ref{fig:solo}), and $\hat{\mathbf{y}}$ completes the right-handed coordinate system.
\begin{figure*}
\centering
   \includegraphics[width=17cm]{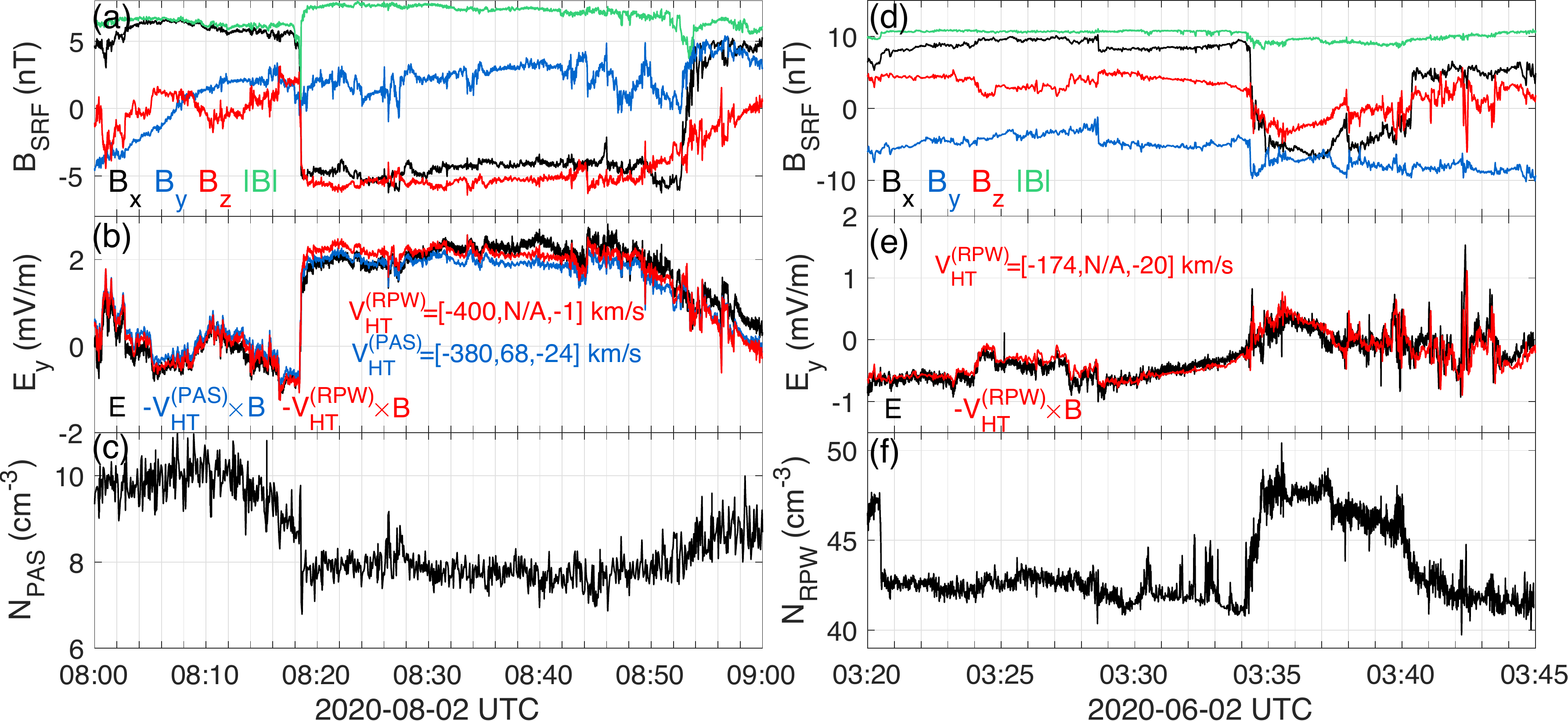}
     \caption{Current sheet crossings at different heliocentric distances (left column 0.75 AU, right column 0.54 AU). (a) Magnetic field in SRF coordinates. (b) $E_y$ (black) and $-(\mathbf{v}_{\text{HT}}\times \mathbf{B})_y$ with $\mathbf{v}_{\text{HT}}$ from PAS (blue) and RPW (red). (c) Plasma density from PAS. (d-f) Same format as (a-c) for a different current sheet where PAS data is unavailable. Plasma density was estimated from the spacecraft potential~\citep[][this issue]{Khotyaintsev2021}.}
     \label{fig:cs}
\end{figure*}

As the main focus of this study is on the DC electric field $\mathbf{E}$, we briefly summarize the method used to calculate and calibrate $\mathbf{E}$, while referring to~\citet{Maksimovic2020} for more details about the RPW instrument. The electric field is measured using three cylindrical probes lying in the YZ plane as shown in Fig.~\ref{fig:solo}. RPW measures the probe-to-spacecraft potential for the three probes, $V_1, V_2$, and $V_3$, respectively. Since probes 2 and 3 are symmetrically positioned along y with respect to both spacecraft and solar panels, spacecraft effects are mostly canceled out when computing $E_y=(V_3-V_2)/L_{23}$, where $L_{23}$ is the nominal effective antenna length (6.99 m), that is, the spatial distance between the measurements of $V_2$ and $V_3$. This, however, is not the case when we compute $E_z=(V_{23}-V_1)/L_{123}$, where $V_{23}=(V_2+V_3)/2$ is the mean potential of probes 2 and 3, and $L_{123}$ is the nominal effective antenna length (6.97 m) in the z-direction, since probe 1 is placed in a different potential environment compared to probes 2 and 3. As a result, $E_z$ is not as well behaved as $E_y$, and tends to be more noisy. Consequently, throughout this investigation we will only use $E_y$, which is sufficient for the purposes of this study.

In the ideal case $E_y$ can be directly computed as $E_y=(V_3-V_2)/L_{23}$. In reality however, the probe signals may have instrumental offsets which we need to compensate for. The source of the offsets can for example be a difference in characteristics of the individual probes, asymmetries in the electrostatic potential of the spacecraft and solar panels, and plasma wakes. We account for these by taking a large window (several hours) of data and fitting $V_2$ against $V_3$ linearly to find a slope $k_{23}$, and offset $d_{23}$. Typically, $k_{23}$ is steady and very close to 1, and we set it to be identically 1 throughout this study. The offset is more variable with a typical value of $\sim d_{23}=0.1$ V. We therefore correct $V_2$ for the offset $V_2'=V_2-d_{23}$. By this correction we implicitly  assume $E_y=0$ on the time-scale of the window. One additional complication we occasionally need to account for is that we sometimes find an unphysical correlation between $V_2'-V_3$ and $V_2'+V_3$, that is, a correlation between $E_y$ and the spacecraft potential. When this correlation is strong and the absolute value of the correlation coefficient between the two terms exceeds $0.7$, we add an additional step to the calibration to remove this common mode signal. This step is to fit $V_2'-V_3$ linearly against $V_2'+V_3$ yielding an offset $\Gamma_0$ and a slope $\Gamma_1$, which we adjust for symmetrically as $V_2^*=V_2'-[\Gamma_0+\Gamma_1(V_2'+V_3)]/2$, $V_3^*=V_3+[\Gamma_0+\Gamma_1(V_2'+V_3)]/2$. We can then finally compute $E_y$ as $E_y=(V_3^*-V_2^*)/L_{23}$.

\section{Current sheets and deHoffmann-Teller analysis}
\label{sec:cs}
To validate $E_y$, we make use of the plasma property that on large scales the plasma is in the ideal MHD regime, meaning $\mathbf{E}=-\mathbf{v}\times\mathbf{B}$. Since the solar wind velocity is relatively steady if compared to the magnetic field which changes significantly in both magnitude and direction, variations in $\mathbf{E}$ correspond primarily to variations in $\mathbf{B}$. Because of this, current sheet crossings serve as an excellent opportunity to test the DC $\mathbf{E}$ measurements using deHoffmann-Teller (HT) analysis. In this study we calculate $\mathbf{v}_\text{HT}$ in two ways. First, following the theory discussed by~\citet{khrabrov1998} and~\citet{paschmann2008}, dropping the contribution of $E_x, E_z, B_y$ and $v_y$ results in
\begin{equation}
\label{eq:EyHT}
\begin{pmatrix}
\langle E_yB_z\rangle\\
-\langle E_yB_x\rangle
\end{pmatrix}
=
\begin{pmatrix}
    \langle B_z^2\rangle & -\langle B_xB_z\rangle\\
    -\langle B_xB_z\rangle & \langle B_x^2\rangle
\end{pmatrix}
\begin{pmatrix}
v_\text{HT,x}\\
v_\text{HT,z}
\end{pmatrix},
\end{equation}
where $\langle\cdot\rangle$ denotes averaging. Solving for $\mathbf{v}_\text{HT}$, we find the frame in which $E_y=0$.
Evident from Eq.~(\ref{eq:EyHT}) is the fact that we are unable to determine $v_{\text{HT},y}$, and $\mathbf{v}_\text{HT}$ is not unique in that respect. However, as the solar wind is mainly expanding radially from the Sun so that $|v_x|\gg|v_y|$, and the observed $v_y$ is mainly due to orbital motion of the spacecraft, which is rarely of practical importance. Second, when PAS data is available, we use the 3D ion velocity vectors and Eq.~(9.10) from~\citet{khrabrov1998} to obtain an estimate of $\mathbf{v}_\text{HT}$ which does not involve $\mathbf{E}$, and which can be compared to the results of Eq.~(\ref{eq:EyHT}).

In Fig.~\ref{fig:cs} we present two examples of current sheets identified as sharp rotations in at least one component of $\mathbf{B}$, where we apply the HT analysis. For the first current sheet (observed at 0.75 AU) PAS data is available, and we use both methods to calculate $\mathbf{v}_\text{HT}$. We show the results in Fig.~\ref{fig:cs}b, where we plot $E_y$ (black) together with $-(\mathbf{v}_\text{HT}\times\mathbf{B})_y$, where $\mathbf{v}_\text{HT}$ was obtained using RPW data with Eq.~(\ref{eq:EyHT}) (red), as well as PAS data using Eq.~(9.10) from~\citet{khrabrov1998} (blue). The two $\mathbf{v}_\text{HT}$ estimates are in good agreement, $\mathbf{v}^{(\text{RPW})}_\text{HT}=[-400,\text{N/A},-1]$, $\mathbf{v}^{(\text{PAS})}_\text{HT}=[-380,68,-24]$.  There is an excellent agreement between  $E_y$ and $-\mathbf{v}_\text{HT}\times\mathbf{B}$, indicating that the HT frame exists, and that $E_y$ is accurately measured on both large and small scales.
The obtained HT velocities are close to the solar wind velocity measured by PAS, $\mathbf{v}=[-370,64,-18]$ km/s, which shows that the HT method indeed can be used to estimate the solar wind speed. For comparison, the average Alfv\'en speed in this interval is 35 km/s.
The second current sheet (Figs.~\ref{fig:cs}d-f) was observed closer to the Sun, at 0.54 AU, and there was no particle data from PAS to compare with. Using the spacecraft potential and plasma frequency measured by RPW we estimate the plasma density~\citep[][this issue]{Khotyaintsev2021} shown in Fig.~\ref{fig:cs}f, and find an average Alfv\'en speed of 34 km/s. The results of the HT analysis in Fig.~\ref{fig:cs}e shows a good agreement between $E_y$ and $-(\mathbf{v}_\text{HT}\times\mathbf{B})_y$ suggesting again that the HT frame exists and that $E_y$ accurately follows the changes in $\mathbf{B}$. However, we note the low value of $\mathbf{v}^{(\text{RPW})}_{\text{HT}}=[-174,\text{N/A},-20]$ km/s compared to the typical solar wind velocity, which suggests that the magnitude of $E_y$ is underestimated. This is likely related to the effective antenna length $L_\text{eff}$, being shorter than the nominal length $L_{23}=6.99$ m used in the calibration leading to both $E_y$ and $v_\text{HT}$ being underestimated. 

The first current sheet (Fig.~\ref{fig:cs}a-c) had a small magnetic field component normal to the current sheet $B_n\approx0.2$ nT, $B_n/|\mathbf{B}|\approx0.04$, obtained using minimum variance analysis~\citep{Sonnerup1967}, and the plasma data showed no signatures of reconnection. The second current sheet (Fig.~\ref{fig:cs}d-f) had a much larger normal component $B_n\approx1.5$ nT, $B_n/|\mathbf{B}|\approx0.15$, but no plasma data were available to compare with. Next, we investigate a current sheet with an intermediate normal magnetic field component $B_n\approx-0.3$ nT, $B_n/|\mathbf{B}|\approx0.09$, where plasma data, which shows several signatures of magnetic reconnection, were available.

\begin{figure}
  \resizebox{\hsize}{!}{\includegraphics{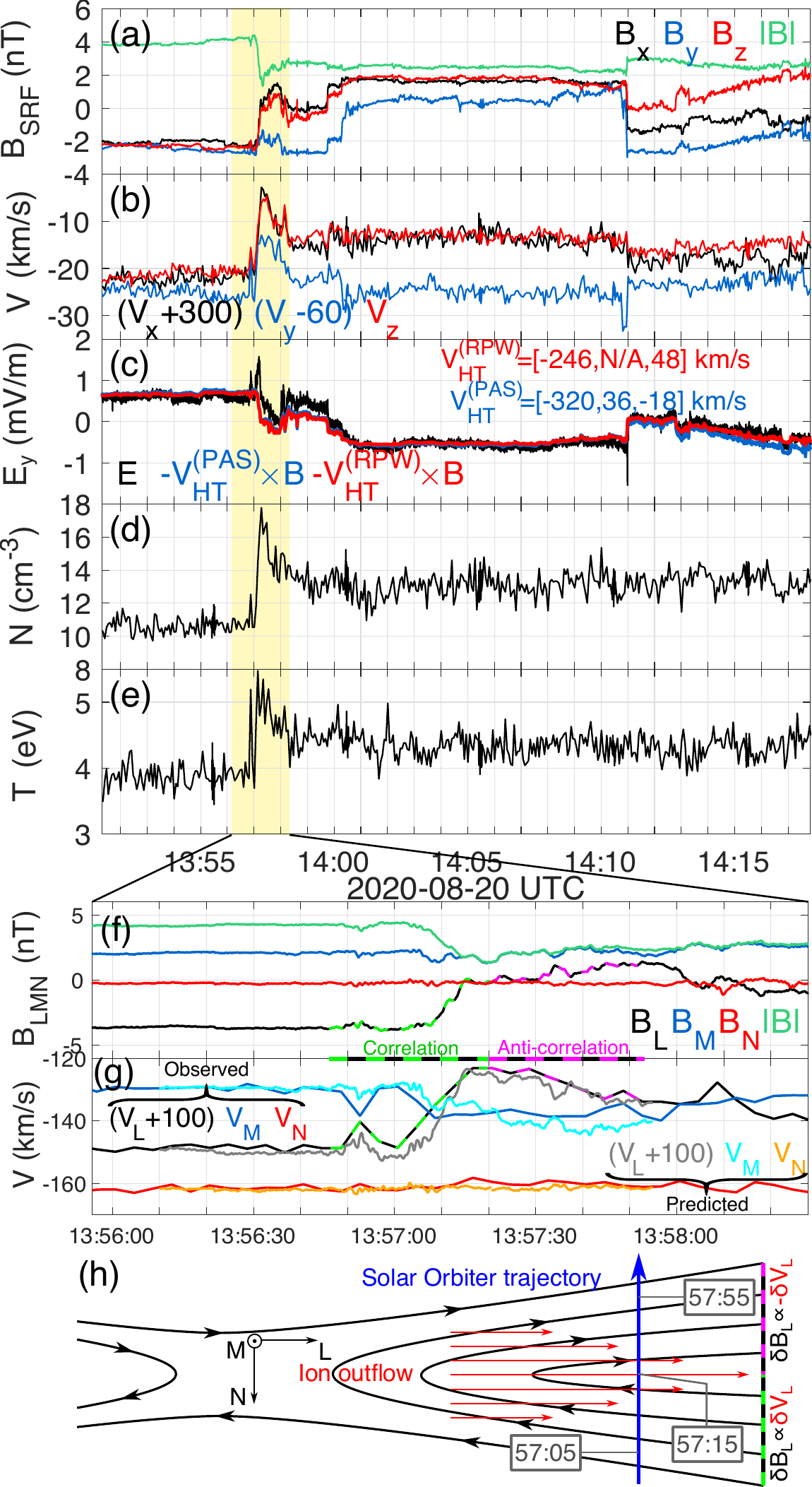}}
  \caption{Observations of a reconnecting current sheet. (a-b) Magnetic field and ion velocity in SRF. (c) $E_y$ and $-(\mathbf{v}_{\text{HT}}\times \mathbf{B})_y$ with $\mathbf{v}_{\text{HT}}$ from PAS (blue) and RPW (red). (d-e) Ion density and temperature from PAS. (f) Magnetic field in LMN coordinates at the reconnection jet. (g) Observed (black, blue, red) and predicted (green, cyan, orange) plasma velocities in LMN coordinates. The green and magenta dashed horizontal bars highlight where $B_L$ and $V_L$ are correlated and anti-correlated, respectively. (h) Simplified sketch of the reconnection event illustrating the expected correlation/anti-correlation between $B_L$ and $V_L$. }
  \label{fig:reconnection}
\end{figure}

On August 20, 2020 Solar Orbiter observed several sharp $\mathbf{B}$ changes during which  $B_x$ changed its sign (i.e. radial polarity changes) often concurrent with significant reductions in magnetic field magnitude, which occasionally reached values as low as $0.1$ nT. The Magnetic Connectivity Tool~\citep{Rouillard2020} shows that Solar Orbiter was very close to the heliospheric current sheet (HCS) during this time, suggesting that these observations were likely multiple crossings of the HCS. In Fig.~\ref{fig:reconnection} we present an overview of two such crossings showing signatures of ongoing magnetic reconnection. Applying our HT analysis on the whole interval (Fig.~\ref{fig:reconnection}c), we again find a good agreement between $E_y$ and $-(\mathbf{v}_{\text{HT}}\times\mathbf{B})_y$, concluding that a HT-frame exists. The difference between $\mathbf{v}^{(\text{RPW})}_\text{HT}=[-246,\text{N/A},48]$ km/s and, $\mathbf{v}^{(\text{PAS})}_\text{HT}=[-320,36,-18]$ km/s, indicates that $|E_y|$ is underestimated by about $20\%$. We again find that the average Alfv\'en speed of 22 km/s is much slower than the solar wind speed, and the effect of Alfv\'enic fluctuations are negligible. During the first crossing at 13:57 UT, highlighted in the figure there is a clear peak in the plasma velocity (Fig.~\ref{fig:reconnection}b) which is concurrent with an increase in density (Fig.~\ref{fig:reconnection}d) and ion temperature (Fig.~\ref{fig:reconnection}e). These are signatures of a reconnection outflow. In order to investigate whether this observation is consistent with ongoing magnetic reconnection we use minimum variance analysis on $\mathbf{B}$ to determine the local current sheet coordinate system~\citep{Sonnerup1967}. We limit the minimum variance analysis to the first small-scale current sheet highlighted in yellow, and find L=[0.75,0.17,0.64], M=[0.31,-0.94,-0.11], N=[0.58,0.27,-0.76], where N is normal to the current sheet, L is in the direction of the reconnecting $\mathbf{B}$ component, and M points out of the reconnection plane. In Figs.~\ref{fig:reconnection}f,g we show a zoom-in on the jet, and present $\mathbf{B}$ and $\mathbf{v}$ in LMN coordinates. One important feature in the data is the changing correlation between $B_L$ and $v_L$, highlighted by the green and magenta dashed lines. During the leading part of the crossing, $B_L$ and $v_L$ are correlated (green dashes), while in the trailing part they are anti-correlated (magenta dashes), consistent with the sketch in Fig.~\ref{fig:reconnection}h. This signature corresponds to the two Alfv\'en waves propagating away from the reconnection site, parallel and anti-parallel to $\mathbf{B}$~\citep[e.g.][]{Gosling2005,Lavraud2009,Phan2020, Froment2021}, and distinguishes it from Alfv\'enic structures such as switchbacks, characterized by a single correlation.
We show the results from a more quantitative test of the magnetic reconnection hypothesis, a Wal\'en test, in Fig.~\ref{fig:reconnection}g, where the overplotted grey, cyan, and orange curves are the predicted velocity components when crossing a rotational discontinuity in an isotropic plasma~\citep{Hudson1970}
\begin{equation}
    \mathbf{v}_\text{predicted}=\mathbf{v}_\text{ref}\pm\left(\frac{\mathbf{B}}{\sqrt{\mu_0\rho}}-\frac{\mathbf{B}_\text{ref}}{\sqrt{\mu_0\rho_\text{ref}}}\right),
\end{equation}
where $\rho$ is the ion mass density, and the subscript 'ref' denotes reference values selected on either side of the jet. The $+$ solution is taken for the leading edge, and $-$ for the trailing edge, and the corresponding time for the reference values are 13:56:10 and 13:57:55 UT. The good agreement between the prediction and observation indicates that Solar Orbiter crossed a rotational discontinuity. Performing the same analysis on similar crossings later in the day (e.g. 16:44:00 UT) yields similar results.

In summary, by applying our HT-method on Solar Orbiter data from crossings of current sheets, some of which are undergoing magnetic reconnection, we find that the HT frames exist, and $E_y$ well captures changes in $\mathbf{B}$. However, due to variations in the effective antenna length, the magnitude of $E_y$, and consequently also $|\mathbf{v}_\text{HT}|$, is not always accurate. We address this issue in the following section.

\section{Solar wind velocity and the effective antenna length}
\label{sec:sw}

As shown in Sect.~\ref{sec:cs}, the shape of $E_y$ is often well represented by $-(\mathbf{v}_\text{HT}\times\mathbf{B})_y$, while the magnitude $|E_y|$, and subsequently also $\mathbf{v}_\text{HT}$, can be off by a scaling factor, likely related to the effective antenna length, $L_\text{eff}$. Understanding how $L_\text{eff}$ varies with plasma conditions is therefore essential for our method to give as good velocity estimates as possible. In the following, we compare plasma velocities from PAS with the estimates from Eq.~(\ref{eq:EyHT}) to model $L_\text{eff}$.

Our procedure is as follows. When PAS, MAG, and RPW data are all available (discrete intervals between the end of May through October 2020), we take a $\pm4$ hour interval around a given time $t_0$. In this interval we compute $E_y$ according to the calibration method discussed in Sect.~\ref{sec:cal}, and band-pass filter both $E_y$ and $\mathbf{B}$ between $0.139$ mHz (2 hour period) and $0.3$ Hz to get rid of potential problems associated with possible drifts of the probe offsets and any high frequency noise. Applying Eq.~(\ref{eq:EyHT}) on the data within $\pm30$ minutes from $t_0$, we finally obtain $\mathbf{v}_{\text{HT}}$. We then take a step of $10$ minutes, $t_0\rightarrow t_0+10[\text{minutes}]$, and repeat this process for all data to obtain a dataset of estimated solar wind velocities.

\begin{figure}
\centering
\resizebox{\hsize}{!}{\includegraphics{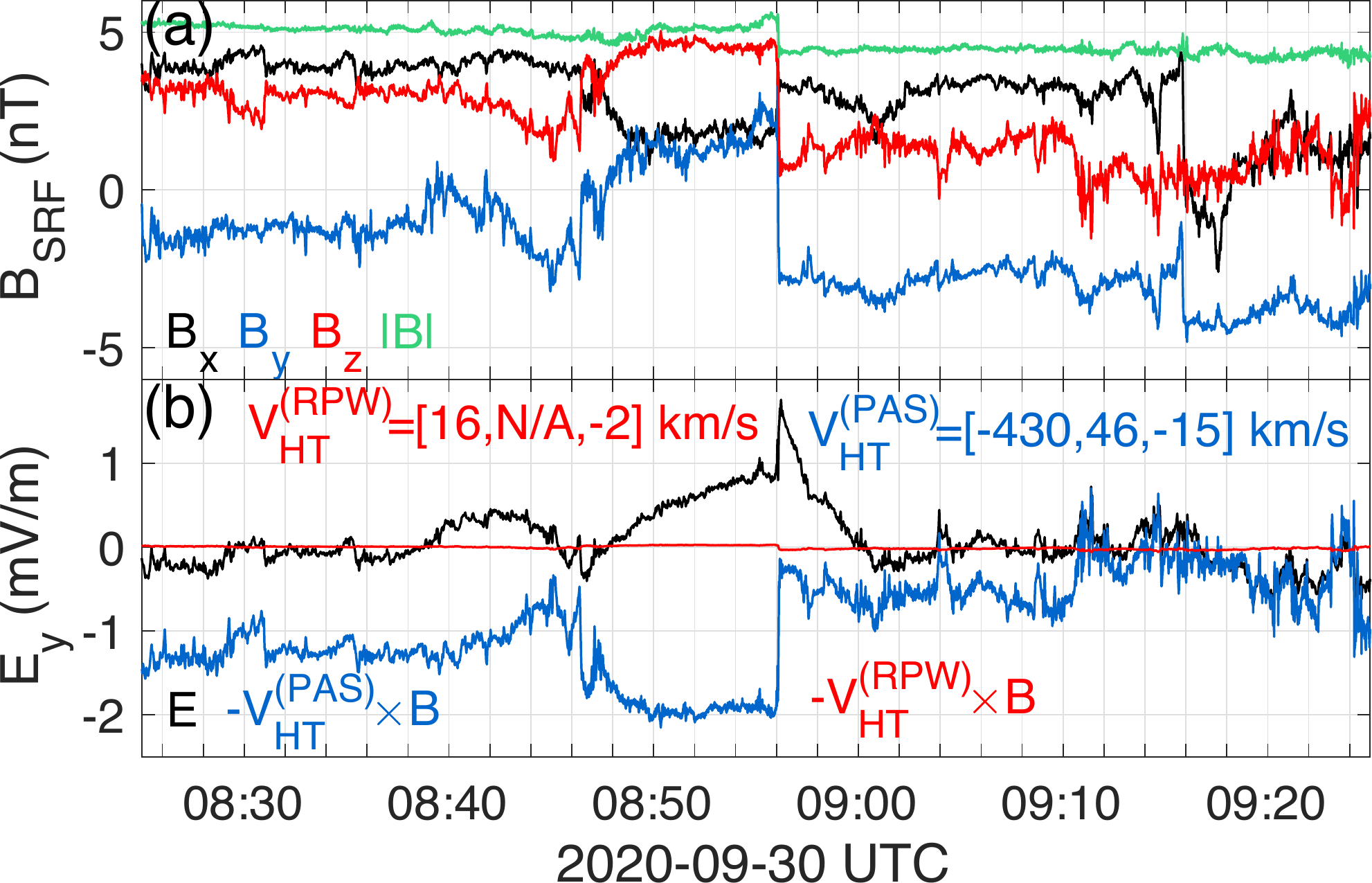}}
\caption{Example of failed HT analysis due to bad $E_y$ data. Same format as Fig.~\ref{fig:cs}a-b.}
     \label{fig:HTbad}
\end{figure}

There are a few ways to quantify the quality of the obtained $\mathbf{v}_{\text{HT}}$. The most straightforward ones are the correlation coefficient and the inclination of the linear slope between $E_y$ and $-(\mathbf{v}_\text{HT}\times\mathbf{B})_y$. For the data to be considered of good quality, we require that the correlation coefficient is larger than $0.9$, and the inclination is between $0.95$ and $1.05$. In Fig.~\ref{fig:HTbad} we show an example of failed HT analysis where clearly $E_y\neq-(\mathbf{v}_\text{HT}\times\mathbf{B})_y$. If the $E_y$ data were good, the measured $E_y$ (black) should have been in good agreement with $-(\mathbf{v}_\text{HT}^\text{(RPW)}\times\mathbf{B})_y$ (red), and similar in shape (but potentially have a different magnitude due to $L_\text{eff}$) to $-(\mathbf{v}_\text{HT}^\text{(PAS)}\times\mathbf{B})_y$ (blue). The cause of this behavior has yet to be identified, but by using the above mentioned quality measures we can discard intervals such as this one, for which the absolute value of the correlation coefficient is well below $0.9$.

\begin{figure}
\centering
\resizebox{\hsize}{!}{\includegraphics{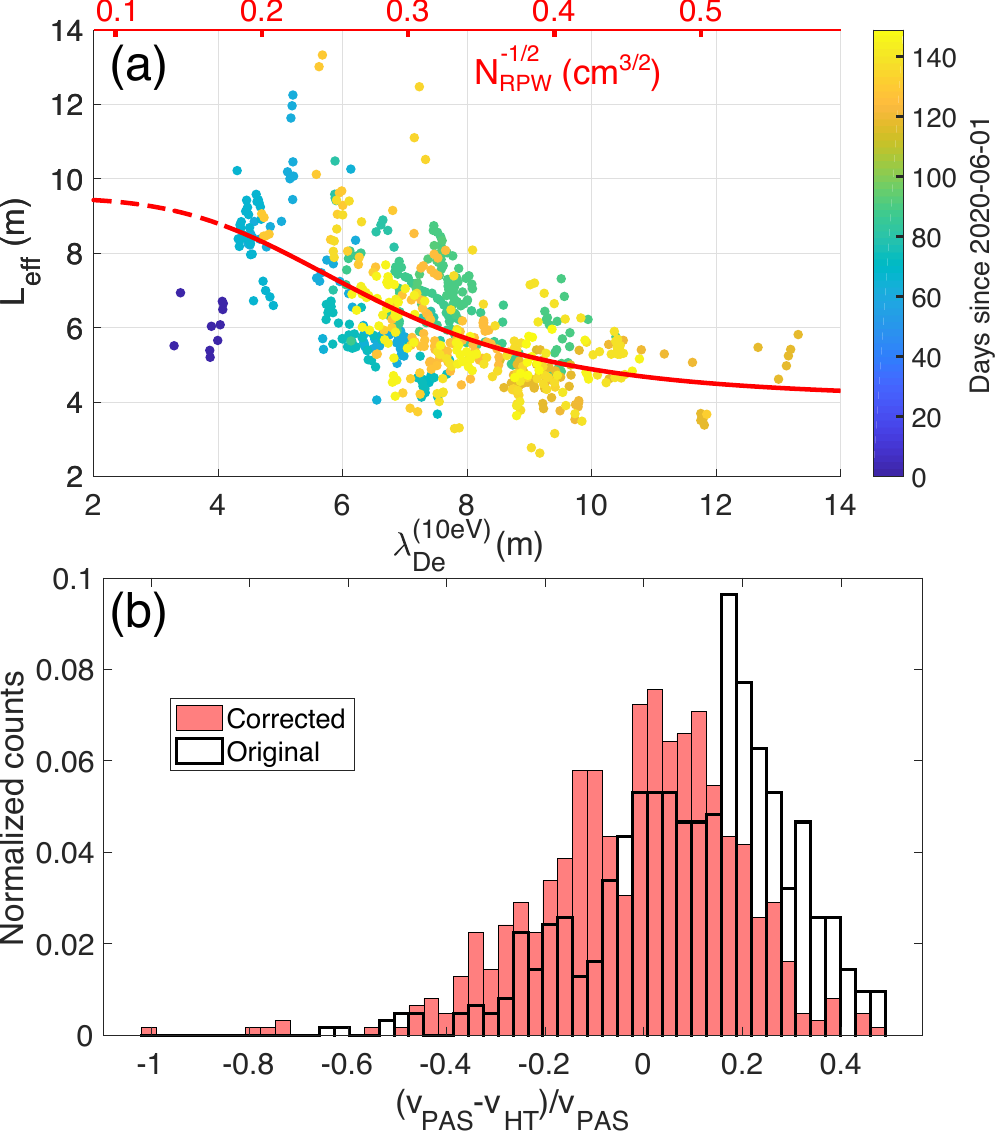}}     \caption{Modeling the effective antenna length. (a) Effective antenna length as a function of $\lambda_{De}$ using densities from RPW from solar wind analysis. Color-coded is the time of each observation in units of days since 2020-06-01. The solid red curve shows a fit to the data (see text for details), and the dashed curve shows the extrapolation. (b) Relative error of solar wind speed before (white) and after (red) correcting for $L_{\text{eff}}$.}
     \label{fig:stats}
\end{figure}

To estimate $L_\text{eff}$, we assume that $v_x$ measured by PAS is exact, so that $v^{(\text{RPW})}_{\text{HT},x}/v_x=E_{y,\text{measured}}/E_{\text{real}}=L_\text{eff}/L_{23}$, where $E_{y,\text{measured}}$ is calculated using the nominal $L_{23}=6.99$m, and $E_{\text{real}}$ is the physical electric field we would obtain by calculating $E_y$ using $L_\text{eff}$. In Fig.~\ref{fig:stats}a, using the data fulfilling our criteria for good quality, we plot $L_\text{eff}$ against $\lambda_{De}=\sqrt{\varepsilon_0 T_e/(e^2 n_\text{RPW})}$, where $e$ is the elementary charge, and we assume a constant electron temperature $T_e=10$ eV, representative at 1 AU~\citep{Wilson2018}, and $n_\text{RPW}$ is the plasma density measured by RPW~\citep[][this issue]{Khotyaintsev2021}. We see that $L_\text{eff}$ can vary by several meters, which is in stark contrast to other missions such as Cluster, where $L_\text{eff}$ was found to be close to constant~\citep{Khotyaintsev2014}. This difference is likely due to the fact that Solar Orbiter's E-field probes are cylindrical, close to the spacecraft, and in a plasma where the Debye length is comparable to the antenna length, whereas Cluster's probes are spherical and located around 40 meters away from the spacecraft, much shorter than the typical Debye length in the magnetosphere. We also note that the effective antenna length of Parker Solar Probe, which has probes comparable to Solar Orbiter, has been found to be highly variable~\citep{Mozer2020b}.
While the observed range of $L_\text{eff}$ is quite large, the results are reasonable in that $L_\text{eff}<14$m, which is the tip-to-tip distance of the probes as shown in Fig.~\ref{fig:solo}. In addition, we see that for increasing Debye lengths $L_\text{eff}$ tends to decrease, and appears to converge to a steady value. Similarly, for Debye lengths approaching zero, we expect $L_\text{eff}$ to reach a constant value, since any effects from the spacecraft potential will be removed by the Debye shielding, however there is not enough data at small $\lambda_{De}$ to give a conclusive picture of the asymptotic behaviour.
We note that the data from the beginning of June (dark-blue dots) are clustered at a significantly shorter $L_\text{eff}$ compared to later data with similar Debye length. We believe this difference can be attributed to changes in the probe surface and photo-electron yield observed during the first perihelion.

We construct a simple semi-empirical model for $L_\text{eff}$ of the form
\begin{equation}
\label{eq:Leff}
    L_\text{eff}\left(\lambda_{De}\right) = L_\text{eff,min}+\frac{L_\text{eff,max}-L_\text{eff,min}}{1+(\lambda_{De}/L_{\text{antenna}})^4},
\end{equation}
where $L_{\text{antenna}}=6.5$m is the length of each individual RPW antenna, and optimization gives $L_\text{eff,min}=4.1$m $L_\text{eff,max}=9.5$m, which are the asymptotic values for large and short $\lambda_{De}$, respectively. This is of course just one of many possible equations to fit the data to, but we favor it due to its relative simplicity and the fact that $L_\text{eff,min}$ and $L_\text{eff,max}$ are within the limiting values expected by Fig.~\ref{fig:solo} (3.76 and 13.99 m respectively). We plot Eq.~(\ref{eq:Leff}) in Fig.~\ref{fig:stats}a in red.
This model can serve as a first order approximation of $L_\text{eff}$. We use the model to correct the velocities in our dataset which satisfy the criteria for good quality described above, as $v_\text{corrected}=v_{\text{HT},x}L_{23}/L_\text{eff}$. We plot the relative error between $v_{\text{HT},x}$ and $v_x$ before and after correction as histograms in Fig.~\ref{fig:stats}b. The correction leads to a clear improvement, with the corrected data being centered around 0, and $71\%$ of the corrected data has a relative error within $\pm20\%$, and $90\%$ of data within $\pm30\%$. For the original data these numbers are $58\%$ and $81\%$, respectively.

\section{Discussion}

Our results open the door to the possibility of using magnetic and electric field data to estimate the solar wind speed in the absence of plasma measurements. For the method to give reliable results for a set of data, variations in the magnetic field due to current sheets or MHD turbulence as well as an estimate of the effective antenna length are crucial. 

The need for magnetic fluctuations of substantial amplitude limits the time resolution of the computed solar wind velocity. In addition, by applying the HT method on sequential intervals, we often find $v_\text{HT}$ to vary more than we expect the solar wind velocity to do. It is therefore often appropriate to average our velocities on the scale of a few hours. For these reasons we believe our method should primarily be used to estimate the solar wind speed and distinguish between fast and slow solar wind on a large scale. However, the method is very flexible, and could in principle be used to determine the phase speed of structures on a shorter time scale as well.

Our simple model of $L_\text{eff}$ in Eq.~(\ref{eq:Leff}) enables us to estimate $L_\text{eff}$ using density measurements, which can be provided by the spacecraft potential and plasma frequency measured by RPW
~\citep[][this issue]{Khotyaintsev2021}. We note that the current model does not account for any variations in $T_e$ or other effects depending on heliocentric distance. \citet[][this issue]{Kretzschmar2021} have used $\mathbf{E}$ and $\mathbf{B}$ measurements of electromagnetic whistler waves to compute $L_\text{eff}$. Their results are in good agreement with our model. Their method could thus be important in the future to obtain local values of $L_\text{eff}$ even if no plasma data are available.

\section{Conclusions}
In summary, we investigate the quality of the $E_y$-component of Solar Orbiter's low frequency 'DC' electric field by applying deHoffmann-Teller analysis on solar wind current sheets, one of which (likely to be the heliospheric current sheet) shows several signatures of ongoing magnetic reconnection. We conclude that $E_y$ is often accurate to a scaling factor likely related to the effective antenna length.
By applying the deHoffmann-Teller analysis to estimate the solar wind velocity and comparing the results to measured plasma velocities, we are able to model how the effective antenna length depends on the Debye length. By applying this model, we can significantly improve our estimate of the solar wind speed, $90\%$ of measurements having relative errors below $30\%$.

Since we can use the RPW measurements of the spacecraft potential and plasma frequency to estimate the plasma density \citep[][this issue]{Khotyaintsev2021}, our results can be used in the future to measure the solar wind speed with sufficient accuracy to distinguish fast and slow solar wind, even in the absence of plasma measurements.

\begin{acknowledgements}
      We thank the entire Solar Orbiter team and instrument PIs for data access and support. Solar Orbiter data are available at http://soar.esac.esa.int/soar/\#home. This work is supported by the Swedish Research Council, grant 2016-05507, and the Swedish National Space Agency grant 20/136. G.C. is supported by Swedish National Space Agency contract SNSA 144/18, and Swedish Research Council contract VR 2018-03569. E.Y. is supported by the Swedish Civil Contingencies Agency, grant 2016-2102. The Solar Orbiter magnetometer was funded by the UK Space Agency (grant ST/T001062/1). T.H. is supported by STFC grant ST/S000364/1 The Solar Orbiter Solar Wind Analyser (SWA) PAS were designed, created, and are operated under funding provided in numerous contracts from the UK Space Agency (UKSA), the UK Science and Technology Facilities Council (STFC), the Centre National d’Etudes Spatiales (CNES, France), the Centre National de la Recherche Scientifique (CNRS, France), and the Czech contribution to the ESA PRODEX programme. Data analysis was partly performed with the AMDA science analysis system provided by the Centre de Données de la Physique des Plasmas (CDPP) supported by CNRS, CNES, Observatoire de Paris and Universit\'e Paul Sabatier, Toulouse.
\end{acknowledgements}

%
%
 \bibliographystyle{aa} 
\bibliography{biblio} 
\end{document}